\documentclass[a4paper,11pt]{article}
\pdfoutput=1 % if your are submitting a pdflatex (i.e. if you have
\usepackage{jcappub} % for details on the use of the package, please
\usepackage{bbm}
\usepackage{amsfonts}
\usepackage{amsmath}
\usepackage{amsfonts,color}
\usepackage{amssymb,float}
\usepackage[utf8]{inputenc}
\usepackage{enumerate}

\usepackage{color}

 \usepackage{hyperref}
\usepackage{enumitem}

\newcommand{\te}{\tilde{e}}
\newcommand{\tA}{\tilde{A}}
\newcommand{\Lag}{{\mathcal L}}

\newcommand{\he}{\hat{e}}
\newcommand{\hL}{\hat{\Lambda}}
\newcommand{\hg}{\hat{g}}

\newcommand{\mpl}{M_{\rm Pl}}
\newcommand{\TT}{\mathbb{T}}

\newcommand{\be}{\begin{equation}}
\newcommand{\ee}{\end{equation}}

\newcommand{\bea}{\begin{eqnarray}}
\newcommand{\eea}{\end{eqnarray}}

\begin{document}

\title{Non-Linear Obstructions for Consistent New General Relativity}

%\author{Jose Beltran Jimenez}
%\email{jose.beltran@usal.es}
%\affiliation{Departamento de F\'isica Fundamental, Universidad de Salamanca, E-37008 Salamanca, Spain}
%
%
%
%\author{Konstantinos F.	Dialektopoulos}
%\email{dialektopoulos@na.infn.it}
%\affiliation{Center for Gravitation and Cosmology, College of Physical Science and Technology, Yangzhou University, Yangzhou 225009, China}
%\affiliation{Aristotle University of Thessaloniki, Thessaloniki 54124, Greece}

\author[a]{Jose Beltr\'an Jim\'enez}
\affiliation[a]{Departamento de F\'isica Fundamental and IUFFyM, Universidad de Salamanca, E-37008 Salamanca, Spain}
\emailAdd{jose.beltran@usal.es}
\author[b,c]{and Konstantinos F.	Dialektopoulos}
\affiliation[b]{Center for Gravitation and Cosmology, College of Physical Science and Technology, Yangzhou University, Yangzhou 225009, China}
\affiliation[c]{Aristotle University of Thessaloniki, Thessaloniki 54124, Greece}
\emailAdd{dialektopoulos@na.infn.it}

\abstract{We revisit the field content and consistency of the New General Relativity family of theories. These theories are constructed in a geometrical framework with a flat and metric-compatible connection, so the affine structure is entirely determined by the torsion. The action is given by a local and parity-preserving quadratic form of the torsion with three free parameters. It is well-known that a special choice of parameters gives an equivalent to General Relativity and that the spectrum of the general linear theory around Minkowski contains an additional 2-form field. It has been suggested that the viability of these theories at linear order requires the 2-form field to feature a gauge symmetry so that it describes a massless Kalb-Ramond field. In this work we revisit these previous results and compute the cubic interactions. We also obtain the decoupling limit of the theories and show that the required gauge symmetry for the 2-form at linear order cannot be extended to higher orders. This signals towards a pathological behaviour of these theories and singles out the equivalent of General Relativity as the only consistent New General Relativity theory with a stable Minkowski background that includes gravity.}
\date{\today}


%\begin{abstract}
%Abstract
%\end{abstract}

\maketitle

\section{Introduction}
General Relativity (GR), devised more than a century ago, is the sector of our standard model that describes gravity as the interaction of matter mediated by a massless spin 2 field, namely the graviton. A direct consequence of its masslessness is the necessity of a gauge symmetry in order to maintain the explicit Lorentz invariance which, in turn, also leads to the need for the equivalence principle in the lowest energy limit \cite{Weinberg:1964ew}. It is precisely the equivalence principle what makes it possible to formulate gravity in geometrical terms. The most widely used geometrical formulation of gravity associates it with the curvature of a spacetime provided with a flat and metric-compatible connection, i.e., the Levi-Civita connection. This formulation deprives the connection of any role and trivialises all the potential dynamics associated to its components. This is the formulation originally incepted by Einstein, who arrived there from a very insightful desire to identify gravity and inertia without any mention to a graviton. Despite the usefulness and success of interpreting gravity as the curvature of spacetime, that has led to enormous advances in understanding the physics of black holes or the cosmological evolution, it is worth bearing in mind that a general geometrical framework allows for a much richer affine structure where the torsion and/or the non-metricity can also be given non-trivial roles. Remarkably, the very same dynamics described by GR can in turn be accommodated within flat geometries, where the curvature vanishes, but gravity can be equivalently interpreted purely in terms of the torsion \cite{Teleparallel,Krssak:2018ywd} or the non-metricity \cite{Nester:1998mp,BeltranJimenez:2017tkd}. Although these three descriptions of GR are equivalent \cite{BeltranJimenez:2019tjy}, they serve as starting points for different paths to modify gravity. 
In this note we will be concerned with the so-called New General Relativity (New GR) theories introduced in \cite{Hayashi:1979qx} that are defined by a local and parity-preserving quadratic form of the torsion. These theories are naturally formulated in a Weitzenb\"ock space and have been extensively explored in the literature. Specifically, solar system tests were already studied in \cite{Hayashi:1979qx}, solutions for highly symmetric configurations have been obtained \cite{NewGRapp}, the validity of the equivalence principle was discussed in \cite{Shirafuji:1996im}, the energy and momentum was studied in \cite{Shirafuji:1997wy} and the propagation of gravitational waves was analyzed in \cite{Hohmann:2018jso}. In the latter, they considered two methods: one perturbative and another one using the Newman-Penrose formalism. Finally, the Hamiltonian analysis of these theories was studied in \cite{Nester:2017wau,Blixt:2019ene,Blixt:2018znp} and a non-local extension was considered in \cite{Koivisto:2018loq}.

If the New GR theories are to be regarded as extensions of GR so that it contains the graviton sector, it is known that the perturbative degrees of freedom (dof's) around Minkowski must consist of the massless spin-2 field and a massless 2-form field, i.e. a  Kalb-Ramond field (see e.g. \cite{OrtinBook}. Crucially, the parameters of the action must be chosen so that the 2-form field features a gauge symmetry in the linear theory necessary to avoid the presence of ghostly modes. A natural question is then if this required symmetry for the quadratic Lagrangian transcends the linear theory and can be maintained or consistently deformed when the interactions are included. Our main interest in this work will be to clarify certain properties of the quadratic action (such as the role of Diffs invariance and the need for a gauge symmetry in the 2-form sector) and, more importantly, to study the properties of the cubic Lagrangian. Whilst most previous studies have focused on the linear sector of New GR, in this work we are concerned about the interactions of the perturbative dof's. In particular, we will show how the gauge symmetry of the quadratic Lagrangian is necessarily broken by the cubic interactions. We will work out the decoupling limit of the theory where the problematic modes can be isolated and, in that limit, we will perform a detailed analysis on the impossibility of having a gauge symmetry at cubic order by showing the non-existence of the would-be Bianchi identities. As we will explain, this loss of the gauge symmetry when the interactions are considered results in a strong coupling problem.

The paper is organised as follows: First, in Sec. \ref{Sec:NGR} we will introduce the required geometrical framework and formalism where New GR is formulated. In Sec. \ref{sec:linearized} we will review some general properties of the linearized theory around Minkowski and explain why an enhanced gauge symmetry is required. In Sect. \ref{sec:cubic} we will consider the cubic interactions. We will compute the cubic Lagrangian and discuss the realisation of Diffeomorphisms at that order. In order to analyse in a cleaner manner the cubic interactions, we introduce an appropriate decoupling limit and obtain the cubic Lagrangian in that limit. For that Lagrangian, we will prove that it is not possible to maintain the gauge symmetries of the quadratic Lagrangian when interactions are included. In Sec. \ref{sec:Palatini} we show the equivalence of the Palatini formulation of the theory with the most widely used one in terms of the vierbeins. Finally, in Sec. \ref{sec:discussion} we summarize our results and discuss them in detail.

\section{New General Relativity}
\label{Sec:NGR}

Before presenting the Lagrangian of New GR, we need to introduce the necessary geometrical framework. The fundamental fields in New GR are the vierbeins\footnote{It is also possible to formulate New GR directly in the metric formalism by employing a Palatini approach. We will stick for the moment to the most extensively used formulation of the theory in terms of the vierbein and we will show the equivalence of the Palatini formulation in Sec. \ref{PalatiniNGR}.} (also called tetrad) $e^{a}{}_{\mu}$ that represent a set of Lorentz frames. Since they are not constrained by any condition, they carry a total of $d^2$ components in $d$ dimensions, i.e., they carry an extra of $d(d-1)/2$ components compared to the metric; those additional components are contained in the antisymmetric part of the vierbein that is given by a $d\times d$ matrix .

The inverse $e_a{}^\mu$ of the vierbein is defined by the equivalent relations:
\begin{equation}
e^{a}{}_{\mu}e_a{}^{\nu} = \delta _{\mu}^{\nu} \quad \text{and} \quad e^a{}_{\mu} e_b{}^{\mu} = \delta ^a_b \,.
\end{equation} 
The spacetime metric is then constructed in terms of the vierbein as
\begin{equation}
g_{\mu\nu} = \eta _{ab}e^a{}_{\mu}e^b{}_{\nu}\,,
\end{equation}
with $\eta_{ab}$ being the flat Minkowski metric. 

New GR is formulated in a spacetime with a flat and metric-compatible connection. The flatness condition requires the connection to be pure gauge or, in other words, it must be given by a general $GL(4,\mathbb{R})$ transformation. The metric compatibility enforces the antisymmetry of the connection. These two conditions are the defining properties of a Weitzenb\"ock  spacetime providing the New GR dwell. 

The Weitzenb\"ock connection is characterised by the torsion which in turn can be fully expressed in terms of the vierbein as\footnote{We implement a vanishing spin connection, i.e. $\omega ^a{}_{b} = 0$ in order to avoid using Lagrange multipliers \cite{Nester:2017wau}.}
\begin{equation}
T^a{}_{\mu\nu} =2\partial _{[\mu}e^a{}_{\nu ]}\,.
\end{equation}
The torsion can be projected by using the co-tetrad fields as
\be
T^\alpha{}_{\mu\nu} =2e_a{}^\alpha\partial _{[\mu}e^a{}_{\nu ]}\,.
\label{projT}
\ee

After briefly introducing the necessary geometrical framework, we can proceed to the construction of the theory. The criteria to built the New GR Lagrangian are:
\begin{itemize}
\item Global Lorentz invariance.

\item Diffeomorphisms invariance.

\item At most quadratic in the torsion.

\item No higher than second order field equations.
\end{itemize}

The only terms that transform as scalars under general coordinate transformations and Lorentz transformations and are quadratic in first partial derivatives of the vierbein are the so-called \textit{Weitzenb\"ock invariants}
\begin{equation}
I_1 = T _{\rho\mu\nu}T^{\rho\mu\nu}\,,\quad I_2 = T _{\rho\mu\nu}T^{\nu\mu\rho}\,,\quad I_3 =  T{}^{\rho}{}_{\mu\rho}T^{\sigma\mu}{}_{\sigma}\,.
\end{equation}
In 4 dimensions one can construct the additional parity-odd invariant\footnote{Strictly speaking, in a general metric-affine geometry there are two independent parity-odd invariants, the other being $I_5=\epsilon^{\mu\nu\rho\sigma} T_{\mu\nu\lambda} T^{\rho\sigma\lambda}$. In our Weitzenb\"ock geometry, they are related via integrations by parts.} $I_4=\epsilon^{\mu\nu\rho\sigma} T^\lambda{}_{\lambda\mu} T_{\nu\rho\sigma}$. We will stick to the parity-preserving section in this work as in the original New GR proposal, so we will not consider parity-breaking terms\footnote{The parity-breaking sector was considered in \cite{MuellerHoissen:1983vc} where it was argued that it could in fact play a crucial role for the well-posedeness of the Cauchy problem for the linearised theory within a certain sub-class of New GR theories. As we will discuss below, we do not think there is really such a problem for the linearised theory.}.

The Lagrangian of New General Relativity is then the sum of the above independent invariants
\begin{equation}\label{LNGR}
\mathcal{L}_{\text{NGR}} =\frac12\mpl^2 \,e\sum _{i=1} ^{3}c_i I_i\,,
\end{equation}
with $e$ being the determinant of the tetrad, $e =\det{e^a{}_{\mu}}$, $\mpl$ is the Planck mass and $c_i$ are three arbitrary dimensionless parameters. There is of course a cosmological constant term that also satisfies our requirements, but we will consider it as part of the matter sector. For the particular case where $c_1 = 1/4,\, c_2 = 1/2,$ and $c_3 = -1,$ the above Lagrangian is the one of the Teleparallel Equivalent of General Relativity (TEGR), i.e. the Einstein-Hilbert Lagrangian up to a boundary term.  A crucial feature of this particular choice of parameters is that it singles out a Lagrangian where the global Lorentz invariance of the general quadratic theory is promoted to a a local Lorentz invariance. This enhanced local invariance precisely allows to have the appropriate number of dof's that describe the massless spin-2 graviton. These features will be shown at the perturbative level below.

\section{The quadratic action around Minkowski}
\label{sec:linearized}

After introducing the general Lagrangian for New GR, we will now proceed to study its properties. The first thing that should be clarified is the number of perturbative degrees of freedom around some relevant background solutions. We will first consider a Minkowski background so that the vierbein is given by
\begin{equation}
e^{a}{}_{\mu} = \delta ^a{}_{\mu} + A^a{}_{\mu}\,.
\end{equation}
with $A^a{}_\mu$ the corresponding perturbation. With this perturbed vierbein, the co-frame field can be expressed as the perturbative series
\be
e_a{}^\alpha=\sum_{n=0}^\infty (-1)^n \delta^\alpha{}_{a_1}A^{a_1}{}_{\alpha_1}\cdots\delta^{\alpha_{n-1}}{}_{a_n}A^{a_n}{}_{\alpha_n}\delta^{\alpha_n}{}_a=\sum_{n=0}^\infty(-1)^n(A^n)_a{}^\alpha.
\ee
We can then compute the co-field at first order as
\begin{equation}
e_a{}^{\mu} = \delta _a{}^{\mu} - \delta _b{}^{\mu}\delta _a{}^{\nu} A^b{}_{\nu} + \mathcal{O}(A^2) = \delta _a{}^{\mu} - A^{\mu}{}_a+\mathcal{O}(A^2)\,\quad \text{where} \quad A_a{}^{\mu} \equiv \delta_b{}^{\mu}\delta _a{}^{\nu}A^b{}_{\nu}\,.
\end{equation}
The metric at first order is given by 
\be
g_{\mu\nu} = \eta_{\mu\nu} + 2\eta_{ab}\delta^{a}{}_{(\mu} A^b{}_{\nu)}  + \mathcal{O}(A^2).
\ee
It is convenient to introduce the field $A_{\mu\nu}\equiv \eta_{ab}\delta^{a}{}_{\mu} A^b{}_{\nu}$, whose inverse relation is $A^a{}_\mu=A_{\alpha\mu}\delta^\alpha{}_b\eta^{ab}$. The metric up to first order can then be written as
\be
g_{\mu\nu} \simeq \eta_{\mu\nu} + h_{\mu\nu} 
\ee
with $h_{\mu\nu} \equiv 2 A_{(\mu\nu)}$. The antisymmetric part $A_{[\mu\nu]}$ will be carried by the 2-form field
\be
b_{\mu\nu} \equiv 2 A_{[\mu\nu]}.
\ee

The torsion can be fully expressed in terms of the vierbein perturbation as
\be
T^a{}_{\mu\nu}=2\partial_{[\mu} A^a{}_{\nu]}
\ee
which is an exact relation for the Minkowski background, i.e., valid at all orders in the perturbations. The projected torsion can then be expressed as
\be
T^\alpha{}_{\mu\nu}=e_a{}^\alpha T^a{}_{\mu\nu}=2\sum_{n=0}^\infty(-1)^n(A^n)_a{}^\alpha \partial_{[\mu} A^a{}_{\nu]}\simeq2\Big(\partial_{[\mu} A^a{}_{\nu]}-A_a{}^\alpha\partial_{[\mu} A^a{}_{\nu]}\Big)
\label{eq:projtorsion1}
\ee
Since the torsion starts at first order, the Weitzenb\"ock invariants start at second order and will have the schematic form
\be
I_i\sim\sum_{n=0}^\infty A^n(\partial A)^2.
\ee
This is also the form that the New GR Lagrangian will feature. The fact that the leading order of Weitzenb\"ock invariants is second order will greatly simplify the computation of the quadratic and cubic Lagrangians below.

\subsection{Symmetries}
We of course have Diff invariance, which can be easily realised in terms of the perturbations directly. We can also see how the action of a Lorentz transformation affects the perturbations. Since the tetrad is a set of four Diff 1-forms, it transforms under an infinitesimal Diff $x^\mu\rightarrow x^\mu+\xi^\mu$ as
\be
\delta_\xi A^a{}_\mu=-\partial_\mu\xi^\alpha \delta^a{}_\alpha.
\ee
where we have only retained the lowest order in the perturbations. For the field $A_{\mu\nu}$ we then have
\be
\delta_\xi A_{\mu\nu}=\delta_\xi \Big(\eta_{ab}\delta^a{}_{\mu} A^b{}_{\nu}\Big)=-\partial_\nu\xi_\mu.
\ee
In terms of $h_{\mu\nu}$ and $b_{\mu\nu}$, the Diffs are realised as
\bea
\label{DiffMinkowski1}
\delta_\xi h_{\mu\nu}&=&-2\partial_{(\mu}\xi_{\nu)},\\
\delta_\xi b_{\mu\nu}&=&2\partial_{[\mu}\xi_{\nu]}.
\label{DiffMinkowski2}
\eea
This is also the symmetry found in \cite{Koivisto:2018loq} for a higher derivative extension of the quadratic theory. As one would expect, the Diffs are realised in $h_{\mu\nu}$ as usual for the metric perturbations. For the New GR Lagrangian we find however that the transformation of the metric perturbation must be accompanied by a transformation of the 2-form field $b_{\mu\nu}$, which also corresponds to the usual gauge symmetry of a massless 2-form, with the same parameter $\xi_\mu$. We will see below that there is a family of theories for which the transformations of $h_{\mu\nu}$ and $b_{\mu\nu}$ are unleashed from each other so that the theory enjoys an additional 4-parameter gauge invariance. This enhanced symmetry will crucially guarantee the healthiness of the linear theory but it will also be the source of the strong coupling problems arising from the cubic interactions.

As usual, the existence of the  gauge symmetry provided by Diffs invariance leads to a set of Bianchi identities. For New GR around the considered Minkowskian trivial background, the Diffs realised according to (\ref{DiffMinkowski1}) and (\ref{DiffMinkowski2}) give rise to the following Bianchi identities for the linear theory:
\be
\partial_\nu\left(\frac{\delta \Lag^{(2)}}{\delta A_{\mu\nu}}\right)=0
\label{eq:BianchiDiffsA}
\ee
or, in terms of $h_{\mu\nu}$ and $b_{\mu\nu}$
\be
\partial_\nu\left(\frac{\delta \Lag^{(2)}}{\delta h_{\mu\nu}}-\frac{\delta \Lag^{(2)}}{\delta b_{\mu\nu}}\right)=0.
\label{eq:BianchiDiffs}
\ee
Let us notice that these Bianchi identities are not specific for New GR nor for the trivial background, but they will be present for any Diff invariant teleparallel theory such as, for instance $f(I_1,I_2,I_3)$ \cite{Hohmann:2018xnb} or the fully general quadratic theory with higher order derivatives discussed in \cite{Koivisto:2018loq} where the corresponding Bianchi identities were explicitly given for the Minkowski background as well.

 In the presence of a source term $\tau^{\mu\nu}$ that can be decomposed into the symmetric and antisymmetric pieces that couple to $h_{\mu\nu}$ and $b_{\mu\nu}$ as
\be
2A_{\mu\nu}\tau^{\mu\nu}=h_{\mu\nu} T^{\mu\nu}+b_{\mu\nu} J^{\mu\nu},
\ee
the Bianchi identities are consistent with the conservation law
\be
\partial_\mu T^{\mu\nu}=\partial_\mu J^{\mu\nu}.
\ee
If the external source only couples to the symmetric component of the perturbed vierbein, we recover the usual conservation of the energy-momentum tensor.

 Let us finally discuss the behaviour under Lorentz transformations. Since the vierbeins form a Lorentz vector, the change under an infinitesimal Lorentz transformation of their perturbations is
\be
\delta_\Omega A^a{}_\mu=\Omega^a{}_b \delta^b{}_\mu
\ee
where $\Omega^a{}_b$ is the Lorentz generator. From this transformation law we can obtain 
\be
\delta_\Omega A_{\mu\nu}=\delta_\Omega\Big(\eta_{ab}\delta^a{}_{\mu} A^b{}_{\nu}\Big)=\eta_{ab}\delta^a_{\mu}\delta^m{}_\nu\Omega^b{}_m\equiv\Omega_{\mu\nu}.
\ee
Since, by definition of the Lorentz generators, $\Omega_{\mu\nu}$ is antisymmetric, we clearly see that  $h_{\mu\nu}$ does not change, while $\delta_\Omega b_{\mu\nu}=2\Omega_{\mu\nu}$. The translation of these transformation properties in terms of Bianchi identities is that the antisymmetric part of the equations of motion of $A_{\mu\nu}$ trivialise if the theory is locally Lorentz invariant. In the language of the fields $h_{\mu\nu}$ and $b_{\mu\nu}$, the latter statement means that the field equations of $b_{\mu\nu}$ should trivialise. This is easy to understand because we see that $b_{\mu\nu}$ could then be gauged away by means of a local Lorentz transformation. On the other hand, let us emphasise that a global Lorentz symmetry is retained by the New GR Lagrangian with arbitrary coefficients, with the corresponding Noether conserved currents.

These are known results in the literature, but it is convenient to reproduce them here in order to provide a more complete analysis of the perturbative degrees of freedom that we perform in the following.

\subsection{Perturbative degrees of freedom}
We will start by reviewing some known results for the linear perturbations around a Minkowski background and will seize the opportunity to make some comments in passing. Bearing in mind that the three Weitzenb\"ock invariants start at second order, the quadratic Lagrangian can be straightforwardly written as 
\begin{align}
\frac{2}{\mpl^2} \Lag^{(2)}_{\rm NGR} = &2 c_1 \left( \partial _{\gamma}A_{\alpha\beta} \partial ^{\gamma}A^{\alpha\beta} - \partial _{\beta} A_{\alpha\gamma} \partial ^{\gamma}A^{\alpha\beta} \right) \nonumber \\
&+ c_2 \left(\partial _{\alpha}A_{\gamma\beta}\partial ^{\gamma} A^{\alpha\beta} + \partial _{\gamma}A_{\beta\alpha}\partial ^{\gamma}A^{\alpha\beta}-2\partial _{\alpha}A_{\beta\gamma}\partial ^{\gamma}A^{\alpha\beta} \right) \nonumber \\
&+ c_3 \left(\partial _{\beta}A^{\gamma}{}_{\gamma}\partial ^{\beta}A^{\alpha}{}_{\alpha}+\partial _{\alpha}A^{\alpha\beta}\partial _{\gamma}A^{\alpha}{}_{\beta} - 2\partial ^{\beta}A^{\alpha}{}_{\alpha}\partial _{\gamma}A^{\gamma}{}_{\beta} \right)\,.
\label{eq:quadraticA}
\end{align}
%
%\be
%\frac{4}{\mpl^2} \Lag^{(2)}_{\rm NGR}=(\partial A)^2
%\ee
It is however useful to express this quadratic Lagrangian in terms of $h_{\mu\nu}$ and $b_{\mu\nu}$ for a better identification of the perturbative dof's:
\bea
\frac{8}{\mpl^2} \Lag^{(2)}_{\rm NGR}&=&\big(2c_1+c_2\big)\partial_\alpha h_{\mu\nu} \partial^\alpha h^{\mu\nu}-(2c_1+c_2-c_3)\partial_\mu h^{\mu\alpha}\partial^\nu h_{\nu\alpha}-2c_3\partial_\mu h\partial_\nu h^{\mu\nu}+c_3 (\partial h)^2\nonumber\\
&&+(2c_1-c_2)\partial_\alpha b_{\mu\nu} \partial^\alpha b^{\mu\nu}-(2c_1-3c_2-c_3)\partial_\mu b^{\mu\alpha}\partial^\nu b_{\nu\alpha}\nonumber\\
&& - 2(2c_1+c_2+c_3)\partial_\mu b^{\mu\alpha}\partial^\nu h_{\nu\alpha},
\label{eq:quadratic}
\eea
with $h=\eta^{\mu\nu} h_{\mu\nu}$. This quadratic Lagrangian contains 10 polarisations for $h_{\mu\nu}$ and $6$ for $b_{\mu\nu}$ making up the 16 components of the perturbed vierbein, but the Diffs gauge symmetry guarantees the presence of four first class constraints that will reduce the number of physical degrees of freedom to a maximum of 8. It may not be obvious the existence of such constraints, but it is straightforward to check that the corresponding Bianchi identities, either in the form \eqref{eq:BianchiDiffsA} applied to \eqref{eq:quadraticA} or (\ref{eq:BianchiDiffs}) applied to \eqref{eq:quadratic}, are identically satisfied for arbitrary parameters $c_i$.

If we want the NGR theory to have anything to do with gravity, the quadratic action (\ref{eq:quadratic}) better contains the massless spin-2 field that is to describe the graviton. As argued in \cite{OrtinBook}, the phenomenological viability of the theory will require the absence of the coupling between $h_{\mu\nu}$ and $b_{\mu\nu}$ so that we should impose $2c_1+c_2+c_3=0$. This choice of parameters turns out to be very special because, not only it decouples the two fields at linear order, but it enhances the gauge symmetry of the quadratic action by introducing an additional symmetry for the two-form field. This can be easily seen by noticing that the kinetic term of the two form can then be written in terms of $H_{\alpha\beta\gamma}H^{\alpha\beta\gamma}$ with $H_{\alpha\beta\gamma}=\partial_{[\alpha} b_{\beta\gamma]}$ the field strength of the two form that is invariant under $\delta_\theta b_{\mu\nu}=2\partial_{[\mu}\theta_{\nu]}$, for an arbitrary 1-form $\theta_\nu$. Notice that this additional gauge symmetry is degenerate for the two-form with the Diffs symmetry so that, effectively, this choice of parameters decouples the Diffs transformations of the symmetric and antisymmetric pieces of the vierbein. In other words, now the Diffs can act on $h_{\mu\nu}$ and $b_{\mu\nu}$ independently and with different parameters. This can be clearly seen by performing a Diff and a gauge transformation simultaneously
\bea
\delta h_{\mu\nu}&=&-2\partial_{(\mu}\xi_{\nu)},\\
\delta b_{\mu\nu}&=&2\partial_{[\mu}\theta_{\nu]}+2\partial_{[\mu}\xi_{\nu]}.
\eea
The change under a Diff of $b_{\mu\nu}$ can be cancelled by a gauge transformation, so there is a diagonal action with $\theta_\mu=\tilde{\theta}_\mu-\xi_\mu$ that decouples the transformations of $b_{\mu\nu}$ and $h_{\mu\nu}$.

Within this class of theories with an enhanced symmetry, the trivialisation of the 2-form sector further requires $2c_1-c_2=0$ (which then also makes $2c_1-3c_2-c_3=0$) that leads to $c_3=-2c_2=-4c_1$ , i.e., the parameters of TEGR, as expected. For these parameters, the Diffs are fully realised with $h_{\mu\nu}$ and we further recover the local Lorentz invariance, since only $b_{\mu\nu}$ transforms under a Lorentz transformation. As discussed above, a consequence of this symmetry is the trivialisation of the antisymmetric piece of the field equations. In our language, the equations for $b_{\mu\nu}$.

It is worth noticing that the mixing term cannot be removed by a field redefinition. The most general field redefinition that we can perform at this order is simply $h_{\mu\nu}\rightarrow h_{\mu\nu}+\lambda h\eta_{\mu\nu}$ for some arbitrary $\lambda\neq-1/4$ to guarantee invertibility, while $b_{\mu\nu}$ remains the same. However, the mixing between $h_{\mu\nu}$ and $b_{\mu\nu}$ is oblivious to this field redefinition because it generates a term $\partial_\mu b^{\mu\alpha}\partial_\alpha h$ that vanishes identically via integration by parts.

If we do not insist on the NGR theory to have anything to do with gravity, we can explore more general parameters choices. For instance, $2c_1+c_2=0$, $c_3=0$ represents a sub-class of the case with $2c_1+c_2+c_3=0$ discussed above where the graviton sector is trivialised so that this theory simply propagates (at linear order) a massless Kalb-Ramond field. This field can in turn be dualised to a scalar field so that there is only one perturbative degree of freedom for this theory. However, the full viability of such a theory would be subject to the properties of the interactions.

Finally, let us comment that the counting of number of physical dof's in the general theory can differ from the counting performed in \cite{Hohmann:2018jso} because their interest is in gravitational waves interacting with matter so they only study the modes contributing to the linearised Riemann curvature. Since this object is gauge-invariant, it is oblivious to any gauge mode, while our counting includes all possible propagating modes, regardless whether they couple to matter in the standard way or not.

%\begin{itemize}

%\item $2c_1+c_2=0$, $c_3=0$. This is a sub-class of the case with $2c_1+c_2+c_3=0$ discussed above where the graviton sector is trivialised so that this theory simply propagates (at quadratic order) a massless Kalb-Rammond field. This field can in turn be dualised to a scalar field so that there is only one perturbative degree of freedom for this theory.

%\item 

%\end{itemize}

\section{Cubic interactions}
\label{sec:cubic}

In the precedent section we have studied the perturbative degrees of freedom for the general Lagrangian of New GR and we have re-derived some known results in the literature. We are now interested in studying the cubic interactions of New GR and, in particular, to elucidate whether the theory with an enhanced symmetry that propagates a massless graviton plus a Kalb-Ramond field, retains that symmetry at the non-linear order. Again, since the Lagrangian is quadratic in the torsion, which is already first order, we only need the perturbation of the vierbeins up to second order to compute the cubic interactions.

The cubic Lagrangian in terms of $A_{\mu\nu}$ is given by:
\begin{align}
\frac{2}{M_{\rm Pl}^2}\mathcal{L}_{\rm NGR}^{(3)} = &-2 A^{\alpha\beta}\Big[ \partial _{\alpha}A^{\gamma\kappa} (2 c_1 \partial _{\beta}A_{\gamma\kappa}+c_2 \partial _{\beta}A_{\kappa\gamma}) + c_3 \partial _{\alpha} A\partial _{\beta}A-c_3 \partial _{\beta}A \partial _{\gamma}A^{\gamma}{}_{\alpha} -c_3 \partial _{\alpha}A\partial_{\gamma}A^{\gamma}{}_{\beta}- \nonumber \\
&- c_3 \partial _{\alpha}A_{\beta}{}^{\gamma}\partial _{\gamma}A + c_3\partial _{\gamma}A\partial ^{\gamma}A_{\beta\alpha} + c_3 \partial _{\gamma}A^{\gamma}{}_{\alpha}\partial _{\kappa}A^{\kappa}{}_{\beta} + c_3 \partial _{\alpha}A_{\beta}{}^{\gamma}\partial _{\kappa}A^{\kappa}{}_{\gamma}- c_3 \partial ^{\gamma}A_{\beta\alpha}\partial _{\kappa}A^{\kappa}{}_{\gamma} - \nonumber \\
&- c_2 \partial _{\alpha}A_{\gamma\kappa}\partial ^{\kappa}A_{\beta}{}^{\gamma} + c_2 \partial _{\alpha} A_{\kappa\gamma}\partial ^{\kappa}A_{\beta}{}^{\gamma} - c_2 \partial _{\gamma} A_{\kappa\alpha}\partial ^{\kappa}A_{\beta}{}^{\gamma} +c_2 \partial _{\kappa} A_{\gamma\alpha}\partial ^{\kappa}A_{\beta}{}^{\gamma} -\nonumber \\
&-2 c_1 \partial _{\beta}A_{\gamma\kappa}\partial ^{\kappa}A^{\gamma}{}_{\alpha} - c_2\partial _{\beta}A_{\kappa\gamma}\partial ^{\kappa}A^{\gamma}{}_{\alpha} + c_2 \partial _{\gamma}A_{\kappa\beta} \partial ^{\kappa}A^{\gamma}{}_{\alpha} + 2c_1 \partial _{\kappa}A_{\gamma\beta}\partial ^{\kappa}A^{\gamma}{}_{\alpha} -\nonumber \\
&- 2c_1 \partial _{\alpha}A_{\gamma\kappa}\partial ^{\kappa}A^{\gamma}{}_{\beta} -c_2\partial _{\alpha}A_{\kappa\gamma}\partial ^{\kappa}A^{\gamma}{}_{\beta}  \Big] + A \Big[ c_3 \partial _{\gamma}A \partial ^{\gamma}A + c_3 \partial _{\beta}A^{\beta\gamma}\partial _{\kappa}A^{\kappa}{}_{\gamma} - \nonumber \\
&-2 c_3 \partial ^{\gamma}A \partial _{\kappa} A^{\kappa}{}_{\gamma} - 2 c_2 \partial _{\beta}A_{\gamma\kappa}\partial ^{\kappa}A^{\beta\gamma} + c_2 \partial _{\beta}A_{\kappa\gamma}\partial ^{\kappa}A^{\beta\gamma} - 2 c_1 \partial _{\gamma}A_{\beta\kappa}\partial ^{\kappa}A^{\beta\gamma} +\nonumber \\
&+2 c_1 \partial _{\kappa} A_{\beta\gamma} \partial ^{\kappa} A^{\beta\gamma} + c_2 \partial _{\kappa} A_{\gamma\beta} \partial ^{\kappa}A^{\beta\gamma}     \Big]\,,
\label{eq:cubicLag}
\end{align}
where $A= A^{\alpha}{}_{\alpha}$ is the trace of the perturbation. The general form of the cubic Lagrangian is very cumbersome so instead of working directly with it, we will study its decoupling limit in order to clarify the possibility of deforming the enhanced gauge symmetry of the quadratic Lagrangian to the cubic order. But before going into that, let us pause a moment to discuss Diffeomorphisms invariance at this order.

\subsection{Diffeomorphisms invariance}
As we have explained above, the New GR Lagrangian is Diffs-invariant and this symmetry will also be exhibited at cubic order. It is instructive to explicitly show the form of the invariance under Diffs up to the cubic Lagrangian. In order to show that, let us recall that we want to realise the Diffs fully by $A^a{}_\mu$, while all the background quantities and coordinates are kept fixed. From the transformation of the vierbein under infinitesimal Diffs
\be
\delta_\xi e^a{}_\mu\equiv \tilde{e}^a{}_\mu(x)-e^a{}_\mu(x)=-\Lag_\xi e^a{}_\mu
\ee
with $\Lag_\xi$ the Lie derivative, we obtain 
\be
\delta_\xi A^a{}_\mu=-\Lag_\xi e^a{}_\mu=-\xi^\lambda\partial_\lambda A^a{}_\mu-\partial_\mu\xi^\lambda \big(\delta^a{}_\lambda+A^a_{\lambda} \big)
\ee
where we have used that $\delta_\xi e^a{}_\mu\equiv \tilde{e}^a{}_\mu(x)-e^a{}_\mu(x)=\tilde{A}^a{}_\mu(x)-A^a{}_\mu(x)=\delta_\xi A^a{}_\mu$. The transformation of $A_{\mu\nu}$ is
\be
\delta_\xi A_{\mu\nu}=\delta_\xi \Big(\eta_{ab}\delta^a{}_{\mu} A^b{}_{\nu}\Big)=-\partial_\nu\xi_\mu-\xi^\lambda\partial_\lambda A_{\mu\nu}-\partial_\nu\xi^\lambda A_{\mu\lambda}.
\ee
From this transformation for $A_{\mu\nu}$, it is straightforward to compute the corresponding Bianchi identities at this order\footnote{This expression can also be obtained by directly expanding at the corresponding order the exact Bianchi identities
\be
\partial_\mu\left(\frac{\delta\Lag_{\rm NGR}}{\delta e^a{}_{\mu}}e^a{}_\lambda\right)-\frac{\delta\Lag_{\rm NGR}}{\delta e^a{}_{\mu}}\partial_\lambda e^a{}_\mu=0.
\ee}:
\be
\partial_\nu\left[\eta_{\mu\lambda}\frac{\delta\Lag^{(3)}}{\delta A_{\mu\nu}}+A_{\mu\lambda}\frac{\delta\Lag^{(2)}}{\delta A_{\mu\nu}}\right]-\partial_\lambda A_{\mu\nu}\frac{\delta\Lag^{(2)}}{\delta A_{\mu\nu}}=0
\ee
We have explicitly checked that these Bianchi identities are satisfied for the New GR Lagrangian with $\Lag^{(2)}$ and $\Lag^{(3)}$ given in \eqref{eq:quadraticA} and \eqref{eq:cubicLag} respectively.

In terms of the fields $h_{\mu\nu}$ and $b_{\mu\nu}$, the transformation can be written as
\bea
\delta_\xi h_{\mu\nu}&=&2\delta_\xi A_{(\mu\nu)}=-2\partial_{(\mu}\xi_{\nu)}-\xi^\lambda\partial_\lambda h_{\mu\nu}-\partial_{(\mu}\xi^\lambda \Big(h_{\nu)\lambda}+b_{\nu)\lambda}\Big),\\
\delta_\xi b_{\mu\nu}&=&2\delta_\xi A_{[\mu\nu]}=2\partial_{[\mu}\xi_{\nu]}-\xi^\lambda\partial_\lambda b_{\mu\nu}+\partial_{[\mu}\xi^\lambda \Big(h_{\nu]\lambda}+b_{\nu]\lambda}\Big).
\eea
At zeroth order in the fields, we of course recover the linear transformation obtained above, but at first order in the fields (which is the necessary order to realise the symmetry up to the cubic Lagrangian) we see that $h_{\mu\nu}$ and $b_{\mu\nu}$ mix in the transformation.

\subsection{St\"uckelberging New GR and its decoupling limit}
In order to gain a better understanding of the New GR theories we will study their decoupling limit. For a proper definition of this limit, it is convenient to introduce the parameters $\epsilon a_1=c_1-\frac14$, $\epsilon a_2=c_2-\frac12$ and $\epsilon a_3=c_3+1$ so that the TEGR is simply given by $\epsilon=0$. In terms of these parameters, the New GR Lagrangian can be written as:
\be
\Lag_{\rm NGR}=\Lag_{\rm TEGR}+\frac12\epsilon\mpl^2 e\sum _{i=1} ^{3}a_i I_i
\ee
The decoupling limit will then be defined as $\mpl\rightarrow \infty$ and $\epsilon\rightarrow0$, while keeping the scale $M^2\equiv\epsilon\mpl^2 $ fixed. The philosophy of this limit is that the strict limit $\epsilon=0$ is not continuous because some dof's disappear due to the appearance of a local Lorentz symmetry, while the decoupling limit is designed so that the number of propagating dof's is maintained, thus providing a continuous limit. In order to manifestly keep track of the dof's we will introduce St\"uckelberg fields in order to restore the local Lorentz symmetry. To that end, we will parameterise the vierbein as 
\be
e^a{}_\mu=\Lambda^a{}_b\,\te^{\,b}{}_\mu
\ee
with $\Lambda^a{}_b$ an arbitrary Lorentz matrix. Under a local Lorentz transformation $L^a{}_b(x)\in SO(3,1)$, we have the transformation laws
\bea
\Lambda^a{}_b&\rightarrow& \Lambda^a{}_c\, (L^{-1})^c{}_b\\
\te^a{}_\mu&\rightarrow& L^a{}_b\,\te^{\,b}{}_\mu
\eea
so that $e^a{}_\mu$ is trivially invariant under a local Lorentz transformation and so is the New GR Lagrangian constructed with it. The Lorentz matrix $\Lambda^a{}_b$ has six St\"uckelberg fields (3 boosts plus 3 spatial rotations). Instead of working with $\Lambda^a{}_b$ directly, it will be convenient to work with its Lie algebra $\Lambda^a{}_b=\big[\exp\omega]^a{}_b$. Since $\Lag_{\rm TEGR}$ possess a local Lorentz symmetry (up to a total derivative) we will have that $\Lag_{\rm TEGR}[e]=\Lag_{\rm TEGR}[\te]$ and all the dependence on the St\"uckelberg fields will come from
\be
\Lag_{\omega}=\frac12 M^2e\sum _{i=1} ^{3}a_i I_i.
\ee
Thus, in order to study the number and nature of the dofs in New GR we can just study the dynamics of the St\"uckelbergs from $\Lag_\omega$.

It is instructive to keep track of the degrees of freedom in this language. In the New GR Lagrangian, the fundamental field is the vierbein $e^a{}_\mu$ with 16 components. However, since we have the Diffs gauge symmetry, there are 4 first class constraints generating the full Diffs and, therefore, only up to $16-4\times2=8$ components can correspond to propagating degrees of freedom. In the case of the TEGR Lagrangian, the additional local Lorentz symmetry reduces the number of propagating dofs to $8-3(\text{boosts})-3(\text{spatial rotations})=2$ as it should. There are other special cases with fewer propagating dofs as we discuss below. From the St\"uckelberg perspective, the counting of dof's goes as follows: Now we have 16 components of the vierbein and 6 St\"uckelberg fields from the general Lorentz matrix. Again we have Diffs that removes 8 dofs, but now we also have the local Lorentz invariance for any choice of the parameters so that we again end up with up to 8 dofs, as it should. In the unitary gauge, the St\"uckelberg fields are completely removed and the analysis is as usual. It is however convenient to use a gauge where the vierbein is parameterised in the usual ADM form with the lapse and the shift being Lagrange multipliers as in the usual case and analyse the extra degrees of freedom from the St\"uckelberg fields.

The Minkowski background considered in the precedent sections was given in the unitary gauge where the St\"uckelberg fields trivialise $\omega^a{}_b=0$ and $e^a{}_\mu=\delta^a{}_\mu$. This configuration can be straightforwardly extended to any gauge with
\be
\te^a{}_\mu=\delta^a{}_\mu+\frac{1}{\mpl}\tA^a{}_\mu
\ee
at all orders, while the St\"uckelberg fields enter through the Lorentz matrix as
\be
\Lambda^a{}_b=\big(e^{\omega/M}\big)^a{}_b\simeq \delta^a{}_b+\frac{1}{M}\omega^a{}_b+\frac{1}{2M^2} \omega^a{}_m\omega^m{}_b
\label{eq:LLTstu}
\ee
up to quadratic order. We can relate the vierbein perturbations in the unitary gauge $A^a{}_\mu$ with $\tA^a{}_\mu$ by means of the following exact relation
\be
A^a{}_\mu=e^a{}_\mu-\delta^a{}_\mu=\frac{1}{\mpl}\Lambda^a{}_b\tA^b{}_\mu+\big(\Lambda-\mathbbm{1}\big)^a{}_b\delta^b{}_\mu.
\label{eq:AtoAt}
\ee
In the above expressions we have already introduced the appropriate scaling behaviours with $\mpl$ and $M$. That these are the appropriate scalings can be easily understood because the quadratic Lagrangian for $\tA$ is entirely determined by $\Lag_{\rm TEGR}$ that scales with $\mpl$, while that for the $\omega$'s comes from $\Lag_{\omega}$ that scales with $M$. More specifically, the New GR Lagrangian at quadratic order will have the schematic form
\be
\Lag_{\rm NGR}\sim (\partial \tA)^2+(\partial\omega)^2+\frac{M}{\mpl}(\partial\omega)(\partial\tA)+\frac{M^2}{\mpl^2}(\partial \tA)^2
\ee
from where it is immediate to see that the introduced scalings are the appropriate ones and that the decoupling limit defined above decouples the propagators of the St\"uckelberg fields from the $\tA$'s without introducing any divergences. Notice that since the pure $\tA$ sector that survives in the decoupling limit comes from the TEGR piece of the Lagrangian, $\tA$ in fact describes the usual massless spin-2 field. The full Lagrangian of New GR including interactions will have the schematic form
\be
\Lag_{\rm NGR}\sim \sum_{n,m}\Big(\frac{\tA}{\mpl}\Big)^n \Big(\frac{\omega}{M}\Big)^m\left[(\partial \tA)^2+(\partial\omega)^2+\frac{M}{\mpl}(\partial\omega)(\partial\tA)+\frac{M^2}{\mpl^2}(\partial \tA)^2\right]
\ee
where it is apparent how the decoupling limit does decouple the St\"uckelberg fields, that then describe the relevant sector. In practice, this means that we can simply neglect the $\tA$-sector and only consider the perturbations generated by the St\"uckelberg fields. We can also see by taking the decoupling limit in the exact relation \eqref{eq:AtoAt} in which case the vierbein perturbation reduces to
\be
A^a{}_\mu=\big(\Lambda-\mathbbm{1}\big)^a{}_b\delta^b{}_\mu,
\ee
or, in terms of $A_{\mu\nu}$, we have
\bea
A_{\mu\nu}&=&\eta_{ab}\delta^a{}_\mu A^b{}_\nu=\eta_{ab}\delta^a{}_\mu\big(\Lambda-\mathbbm{1}\big)^b{}_c\delta^c{}_\nu\nonumber\\
&\simeq&\frac{1}{M}\omega_{\mu\nu}+\frac{1}{2M^2}\omega_{\mu\alpha}\omega^\alpha{}_\nu.
\label{eq:Amnpertdec}
\eea
We see that, at linear order, the St\"ueckelberg field plays the role of $b_{\mu\nu}$ introduced above (up to a factor of 2). However, beyond the linear order we need to consider the higher order corrections in order to guarantee the non-linear realisation of the local Lorentz symmetry with the $\omega$'s.

The vierbein to second order in the decoupling limit can then be written as
\be
e^a{}_{\mu}=\Lambda^a{}_b\delta^b{}_\mu\simeq\left[\delta^a{}_b+\frac{1}{M}\omega^a{}_b+\frac{1}{2M^2} \omega^a{}_m\omega^m{}_b\right]\delta^b{}_\mu
\ee
For the inverse vierbein we use that the inverse of the Lorentz matrix is given by $(\Lambda^{-1})^a{}_b=\eta^{am}\eta_{nb}\Lambda^n{}_m=\Lambda_b{}^a$ so that 
\bea
e_b{}^{\mu}\simeq%\delta^\mu{}_a\left[\delta_b{}^a+\frac{1}{M}\omega_b{}^a+\frac{1}{2M^2} \omega_{bm}\omega^{ma}\right]\nonumber\\&=&
\delta_a{}^\mu\left[\delta^a{}_b-\frac{1}{M}\omega^a{}_b+\frac{1}{2M^2} \omega^a{}_m\omega^m{}_b\right].
\eea
We can then compute the torsion, whose exact expression in the decoupling limit is 
\be
T^a{}_{\mu\nu}=2\partial_{[\mu}\Lambda^a{}_b\delta^b_{\nu]}=2\partial_{[\mu}\big(e^{\omega/M}\big)^a{}_b\;\delta^b_{\nu]}
\ee
and its projected components are
\bea
T^\alpha {}_{\mu\nu}&=&2\delta_c{}^\alpha\Lambda_a{}^c\partial_{[\mu}\Lambda^a{}_b\delta^b_{\nu]}\\
&=&2\delta_c{}^\alpha\big(e^{-\omega/M}\big)^c{}_a\partial_{[\mu}\big(e^{\omega/M}\big)^a{}_b\delta^b_{\nu]}\nonumber\\
&\simeq& \frac{2}{M}\partial_{[\mu}\omega^\alpha{}_{\nu]}-\frac{1}{M^2}\left(\omega^\lambda{}_{[\mu}\partial_{\nu]}\omega^\alpha{}_{\lambda}+\omega^\alpha{}_\rho\partial_{[\mu}\omega^\rho{}_{\nu]}\right)
\label{Torsiondec}
\eea
where we have used the antisymmetry of $\omega_{\mu\nu}$ and the background vierbein $\delta^a_{\mu}$ to trade Lorentz and spacetime indices. Since we are working on a Minkowski background we do not need to distinguish between tangent space and spacetime indices.

Let us also note that, in this decoupling limit, the spacetime metric is
\be
g_{\mu\nu}=e^a{}_\mu e^b{}_\nu\eta_{ab}=\Lambda^a{}_m\Lambda^b{}_n\delta^m{}_\mu \delta^n{}_\nu\eta_{ab}=\eta_{\mu\nu}
\label{eq:gdec}
\ee
at all orders in $\omega$. This is a consequence of the fact that we are focusing on the dynamics of the Lorentz St\"uckelberg fields, which do not contribute to the metric because all the Lorentz-related vierbeins give the same metric. Thus, we will only need the expansion of the torsion \eqref{Torsiondec} to obtain all the interactions in the decoupling limit. 

To end our analysis of the St\"uckelberging and the decoupling limit of New GR, we will discuss the global Lorentz symmetry. In the decoupling limit, the St\"uckelberg fields become invariant under the restored local Lorentz symmetry given in \eqref{eq:LLTstu}. However, the St\"uckelbergs still know about the global symmetry of New GR as can be seen from the form of the projected torsion in \eqref{Torsiondec}, that is invariant under $\Lambda^a{}_b\rightarrow L^a{}_c\Lambda^c{}_d$ for a constant $L^a{}_b$, together with the fact that $g_{\mu\nu}=\eta_{\mu\nu}$ in that limit as shown in \eqref{eq:gdec}. Thus, this symmetry must be present in the sector of the St\"ueckelber fields. As a matter of fact, the St\"ueckelberg fields will play the role of the Goldstone bosons associated to the spontaneously broken global Lorentz symmetry. At the lowest order in the fields, an infinitesimal global Lorentz transformation given by $\ell_{\mu\nu}\in {\mathfrak {so}}(3,1)$ simply amounts to the shift symmetry $\omega_{\mu\nu}\rightarrow \omega_{\mu\nu}+\ell_{\mu\nu}$ whose consequence will be that the quadratic Lagrangian can only contain derivatives of $\omega_{\mu\nu}$. The non-linear realisation of the global Lorentz symmetry at higher orders will result in non-trivial relations between the coefficients of the different orders in the $\omega-$expansion. As we will show below, the coefficient of the cubic interaction is determined by the quadratic Lagrangian through the global Lorentz symmetry.

\subsection{Revisiting the quadratic Lagrangian in the decoupling limit}

The quadratic Lagrangian for the St\"uckelberg field $\omega_{\mu\nu}$ is, up to total derivatives, the following:
\be
 \Lag^{(2)}_\omega=\frac{2a_1-a_2}{2}\partial_\alpha\omega_{\mu\nu}\partial^\alpha\omega^{\mu\nu}-\frac{2a_1-3a_2-a_3}{2}\partial_\alpha\omega^{\alpha\mu}\partial^\beta\omega_{\beta\mu}.
 \label{Eq:omega2}
\ee
At this order we confirm that the Lagrangian for the St\"uckelberg field is the same as that of the pure $b_{\mu\nu}$-sector in \eqref{eq:quadratic} upon the identification $b_{\mu\nu}=2\omega_{\mu\nu}$ so the discussions made for the general quadratic action can be directly translated to the decoupling limit. We will briefly repeat it here in the decoupling limit for completeness. Since this is the Lagrangian for a 2-form with no mass terms, it seems a good idea to require the corresponding gauge symmetry $\omega_{\mu\nu}\rightarrow\omega_{\mu\nu}+2\partial_{[\mu}\theta_{\nu]}$ so that the St\"uckelberg field describes a healthy massless 2-form field. This is indeed the case for the parameters $2a_1+a_2+a_3=0$. It is interesting to note that this gauge symmetry is degenerate with the Diffs realised on $b_{\mu\nu}$, which at this order coincides with $\omega_{\mu\nu}$. This means that the realisation of the Diffs at this order can be fully realised with $h_{\mu\nu}$. As a matter of fact, beyond the decoupling limit, the quadratic action features a mixing between $h$ and $\omega$ whose coefficient precisely vanishes when $2a_1+a_2+a_3=0$, as we saw in \eqref{eq:quadratic}.

For this choice of parameters we then have that, at quadratic order, the theory propagates 3 physical degrees of freedom, namely: the two polarisations of the graviton plus the dof associated to the massless 2-form field. This is however achieved by an appropriate tuning of the coefficients that gives rise to an additional gauge symmetry at quadratic order. The natural question is then if this gauge symmetry can be extended to higher orders, thus guaranteeing that the 2-form field does propagate one additional dof at all orders.

\subsection{Decoupling limit of the cubic interactions}
After recovering the known results for the linear theory from the decoupling limit, let us now analyse the first interacting order, i.e., the cubic Lagrangian. As we have seen, the promising candidate arising from the linear analysis contains a massless spin 2 field and a Kalb-Rammond field. A necessary condition to guarantee this field content was the enhancement of the Diff gauge symmetry where the antisymmetric part of the vierbein or the St\"uckelberg fields in the decoupling limit feature an additional gauge symmetry. The goal now is to study whether this additional gauge symmetry can be maintained at higher orders and, therefore, if the viability of the linear theory is not spoilt by additional dof's arising from the interactions. Again, since the torsion is already first order in $\omega_{\mu\nu}$ and the Lagrangian is quadratic in the torsion, the cubic interaction can be easily computed as
\be
\Lag^{(3)}_\omega=\frac{1}{2M}\omega^{\alpha\beta}\Big[(1-2a_1+3a_2)\partial_\mu\omega_{\alpha\nu}\partial_\beta\omega^{\mu\nu}+(1-a_3)\partial_\beta\omega_{\alpha\mu}\partial_\nu\omega^{\mu\nu}\Big].
\ee
%It is straightforward to see that the first term is just a total derivative by noticing that $\omega^{\alpha\beta}\partial_\mu\omega_{\alpha\beta}=\frac12\partial_\mu(\omega_{\alpha\beta}\omega^{\alpha\beta})$ so that the first term can be written via integration by parts as $\omega_{\alpha\beta}\omega^{\alpha\beta}\partial_\mu\partial_\nu\omega^{\mu\nu}$ that vanishes due to the antisymmetry of $\omega_{\mu\nu}$. 
The second and third terms can be seen to give the same interaction via integration by parts. Thus, adding the total derivative $\frac12(a_3-1)\partial_{\mu}\big(\omega^{\alpha\beta}\omega_\nu{}^\mu\partial_\beta\omega_\alpha{}^\nu\big)$, the cubic Lagrangian in the decoupling limit can be written as
\be
\Lag^{(3)}_\omega=-\frac{1}{2M}(2a_1-3a_2-a_3)\omega^{\alpha\beta}\partial_\mu\omega_{\alpha\nu}\partial_\beta\omega^{\mu\nu}.
\label{eq:cubicStu}
\ee
The same Lagrangian can be obtained by inserting \eqref{eq:Amnpertdec} directly into $\Lag^{(2)}+\Lag^{(3)}$ given in \eqref{eq:quadraticA} and \eqref{eq:cubicLag}, bearing in mind that the quadratic part will also contribute at cubic order. We have explicitly checked that this method indeed reproduces \eqref{eq:cubicStu}. We could have also guessed this cubic Lagrangian from the global Lorentz symmetry of the New GR Lagrangian. Since this global symmetry is non-linearly realised in the St\"uckelberg fields, it will impose non-trivial relations among the coefficients of the quadratic and cubic Lagrangians. In the decoupling limit and up to first order in the St\"uckelbergs, the global Lorentz symmetry is realised as
\be
\delta_\ell\omega^\alpha{}_\beta=\ell^\alpha{}_\beta+\frac{1}{M} \ell^\alpha{}_\lambda\omega^\lambda{}_\beta+\mathcal{O}(\omega^2)
\label{eq:GlobalLstu}
\ee
with $\ell^\alpha{}_\beta$ a constant element of ${\mathfrak {so}}(3,1)$. If we take a generic cubic Lagrangian
\be
\Lag=d_1\partial_\alpha\omega_{\mu\nu}\partial^\alpha\omega^{\mu\nu}+d_2\partial_\alpha\omega^{\alpha\mu}\partial^\beta\omega_{\beta\mu}+\frac{d_3}{M}\omega^{\alpha\beta}\partial_\mu\omega_{\alpha\nu}\partial_\beta\omega^{\mu\nu}
\ee
with arbitrary parameters $d_i$, its change under \eqref{eq:GlobalLstu} is, up to total derivatives,
\be
\delta_\ell\Lag=\frac{d_3-d_2}{M}\ell^{\alpha\beta}\partial_\alpha\omega_{\beta\mu}\partial_\nu\omega^{\mu\nu}
\ee
where we see that the global Lorentz symmetry requires $d_3=d_2$, which is precisely the relation found for \eqref{eq:cubicStu} and \eqref{Eq:omega2}. This consistency check reassures the correctness of our results.

Besides the discussed global Lorentz symmetry respected by New GR, the obtained cubic Lagrangian shows once again the realisation of the local Lorentz symmetry in the TEGR case ($a_i=0$), up to total derivatives. In other words, for $a_i=0$, the cubic interaction of the St\"uckelberg field is just a total derivative. This cubic interaction trivialises for $2a_1-3a_2-a_3=0$ that, in combination with $2a_1+a_2+a_3=0$ gives $a_2=2a_1$, $a_3=-4a_1$, which precisely corresponds to the TEGR (as can be easily seen by computing the corresponding values of $c_i$). Notice that for these parameters, the quadratic action for the St\"uckelberg field \eqref{Eq:omega2} also trivialises. Thus, if we want to maintain the gauge symmetry of the quadratic action for the St\"uckelberg fields without trivialising them, we have to study if the gauge symmetry of the quadratic Lagrangian can be extended to the cubic order. The appearance of the St\"uckelberg without derivatives in the cubic interaction already signals that, if possible at all, the gauge symmetry must be realised as a perturbative series in $\omega_{\mu\nu}$. Therefore, to check whether there is some gauge symmetry, we need to account for the fact that the symmetry at order $\omega^3$ will have the second order contribution from the quadratic action plus the first order of the cubic interaction. We will then consider the general  zeroth order transformation $\delta^{(0)}\omega_{\mu\nu}=2\partial_{[\mu}\theta_{\nu]}$ supplemented by the first order transformation 
\be
\delta^{(1)}\omega_{\mu\nu}=2\alpha_1\omega_{[\mu}{}^{\lambda}\partial_{\nu]}\theta_\lambda+2\alpha_2\partial_\lambda\theta_{[\mu}\omega_{\nu]}{}^\lambda+ \alpha_3 \omega_{\mu\nu}\partial_\lambda\theta^\lambda+2\alpha_4\partial_{[\mu}\omega_{\nu]\lambda}\theta^\lambda+\alpha_5\partial_\lambda\omega^\lambda{}_{[\mu}\theta_{\nu]}+\alpha_6\theta^\lambda\partial_{\lambda}\omega_{\mu\nu}
\label{deltaomega1}
\ee
where $\alpha_i$ are some parameters (of order $1/M$) and we have included up to first order derivatives. The Bianchi identities at this order will be of the form
\be
\partial_\mu\frac{\delta \Lag^{(3)}}{\delta \omega_{\mu\nu}}+\mathcal{D}^{(1)}_\mu\frac{\delta \Lag^{(2)}}{\delta \omega_{\mu\nu}}=0
\label{Bianchi1}
\ee
where $\mathcal{D}^{(1)}_\mu$ is the differential operator associated to (\ref{deltaomega1}). If we take the term with highest number of derivatives from the first piece in (\ref{Bianchi1}) (three in this case), we obtain 
\be
\partial_\mu\frac{\delta \Lag^{(3)}}{\delta \omega_{\mu\nu}}\supset-\frac{1}{4M}(2a_1-3a_2-a_3)\Big(\omega^{\mu\lambda}\partial_\lambda\Box \omega^\nu{}_\mu+\omega^{\mu\lambda}\partial_\rho\partial_\lambda\partial^\nu\omega_\mu{}^\rho\Big).
\label{Bianchi2}
\ee
From this term, one can convince oneself that it is not possible to cancel both terms simultaneously from the second piece in  (\ref{Bianchi1}). More explicitly, we have
\be
\frac{\delta \Lag^{(2)}}{\delta \omega_{\mu\nu}}=-(2a_1-a_2)\Big( \Box\omega^{\mu\nu}+2\partial_\lambda\partial^{[\mu}\omega^{\nu]\lambda}\Big).
\label{highestcontr}
\ee
The only terms that will generate third order derivatives from $\mathcal{D}^{(1)}_\mu\frac{\delta \Lag^{(2)}}{\delta \omega_{\mu\nu}}$ in \eqref{Bianchi1} are those generated by the terms $\alpha_1$, $\alpha_2$ and $\alpha_3$ in \eqref{deltaomega1}. We can note that the term $\alpha_1$ will only contribute a term with third derivatives of the form $\propto \omega_{[\mu}{}^\lambda\partial_{\nu]}\frac{\delta \Lag^{(2)}}{\delta \omega_{\mu\nu}}$ which vanishes by virtue of the zeroth order Bianchi identities. Thus, only $\alpha_2$ and $\alpha_3$ contribute and we find
\be
\mathcal{D}^{(1)}_\mu\frac{\delta \Lag^{(2)}}{\delta \omega_{\mu\nu}}\supset \frac12(2a_1-3a_2-a_3)\omega^{\mu\lambda}\Big[2\alpha_2\partial_\lambda\Box\omega^\nu{}_{\mu}+2(\alpha_2-\alpha_3)\partial^\nu\partial_\rho\partial_\lambda\omega_\mu{}^\rho+\alpha_3\partial^\nu\Box\omega_{\mu\lambda}\Big].
\ee
We can see that there is no choice for $\alpha_2$ and $\alpha_3$ that can eliminate the third order derivatives generated from \eqref{Bianchi2} in the Bianchi identities \eqref{Bianchi1}. Thus, only if the cubic interaction identically vanishes, the gauge symmetry can be maintained. As we have explained above, this precisely corresponds to TEGR with the local Lorentz symmetry that trivialises the $\omega_{\mu\nu}$-sector. Thus, we have found that the gauge symmetry that makes the theories propagate a massless Kalb-Ramond field and the usual graviton at linear order cannot be extended to higher order. The breaking of this symmetry by the interactions implies the appearance of additional degrees of freedom that will jeopardise the stability of these theories.

\section{Palatini New GR}\label{PalatiniNGR}
\label{sec:Palatini}

For the sake of completeness, we will show how the above results obtained in the most common formulation of New GR in terms of the vierbeins can be recovered within its Palatini formulation. We will closely follow \cite{BeltranJimenez:2018vdo}. The starting point is the very same action determined by a general quadratic form of the torsion $T^\alpha{}_{\mu\nu}=2\Gamma^\alpha{}_{[\mu\nu]}$, but now directly working with the affine connection $\Gamma^\alpha{}_{\mu\nu}$ (so we do not need to go through the veirbein formulation). The flat and metric-compatibility conditions are imposed by adding suitable Lagrange multipliers that enforce the required constraints. Thus, the Lagrangian will be expressed as
\be
\Lag_{\rm Palatini}=\frac12\mpl^2\sqrt{-g}\, \TT+\lambda_\alpha{}^{\beta\mu\nu} R^\alpha{}_{\beta\mu\nu}+\lambda^\alpha{}_{\mu\nu}\nabla_\alpha g^{\mu\nu}.
\ee
where we have introduced the general quadratic form of the torsion 
\be
\TT\equiv c_1 T_{\alpha\mu\nu}T^{\alpha\mu\nu}+c_2T_{\alpha\mu\nu}T^{\mu\alpha\nu}+c_3T_\alpha T^\alpha\,,
\ee
with $T_\mu=T^\alpha{}_{\mu\alpha}$. That this Lagrangian with $c_1=1/4$, $c_2=1/2$ and $c_3=-1$ is equivalent to GR can be easily understood by using the relation between the Ricci scalar of a general connection and that of the Levi-Civita connection of the metric:
\be
R(\Gamma)=R(g)+\frac14 T_{\alpha\mu\nu}T^{\alpha\mu\nu}+\frac12 T_{\alpha\mu\nu}T^{\mu\alpha\nu}-T_\alpha T^\alpha+\frac{2}{\sqrt{-g}}\partial_\mu\Big(\sqrt{-g} T^\mu\Big)
\ee
that holds for a metric-compatible connection. We then see that, for a flat connection with $R(\Gamma)=0$, the Einstein-Hilbert term differs from that of TEGR by a total derivative.

In order to obtain the action for the perturbative dof's we first need to solve the constraints. The Lagrange multiplier $\lambda^\alpha{}_{\mu\nu}$ imposes the flat constraint 
\be
R^\alpha{}_{\beta\mu\nu}(\Gamma)=0.
\ee
Since this simply states that the curvature vanishes, the connection must be pure gauge or, equivalently, it must differ from the trivial connection by a general linear transformation so it must take the form
\be
\Gamma^\alpha{}_{\mu\nu}=(\Lambda^{-1})^\alpha{}_\lambda\partial_\mu \Lambda^\lambda{}_\nu\,
\label{eq:inertialconnection}
\ee
with $\Lambda^\alpha{}_\beta$ an element of $GL(4,\mathbb{R})$, i.e., an arbitrary non-degenerate $4\times4$ matrix\footnote{Let us stress for clarity that this $\Lambda$ is different from the Lorentz matrix used in precedent sections. Also, as a technical point, since we want the theory to be physically sensible as a perturbative expansion around a trivially connected spacetime with $\Lambda=\mathbbm{1}$, strictly speaking $\Lambda$ must belong to the component of $GL(4,\mathbb{R})$ connected with the identity, i.e., $\det \Lambda>0$.}. This connection is sometimes referred to as a purely inertial connection. In terms of this inertial connection, the torsion can be expressed as
\be
T^\alpha{}_{\mu\nu}=2(\Lambda^{-1})^\alpha{}_\lambda\partial_{[\mu} \Lambda^\lambda{}_{\nu]}\,.
\label{TLAmbda}
\ee
Let us notice the resemblance of this expression with \eqref{projT}. On the other hand, the constraint enforced by $\lambda^\alpha{}_{\mu\nu}$ leads to
\be
2(\Lambda^{-1})^\lambda{}_\kappa\partial_\alpha\Lambda^\kappa{}_{(\mu} g_{\nu)\lambda} =\partial_\alpha g_{\mu\nu}\,. 
\label{eq:Lambdag}
\ee
that relates the metric to $\Lambda^\alpha{}_\beta$. Thus, the fundamental field in this formulation is $\Lambda^\alpha{}_\beta$ and the metric must be expressed in terms of this element by solving the above equation. For the Minkowski background solution we will have
\be
\Lambda^\alpha{}_\beta=\delta^\alpha{}_\beta+\lambda^\alpha{}_\beta
\ee
where $\lambda^\alpha{}_\beta$ is the perturbation. We then have
\be
(\Lambda^{-1})^\alpha{}_\beta=\sum_{n=0}^\infty (-1)^n(\lambda^n)^\alpha{}_\beta\simeq\delta^\alpha{}_\beta-\lambda^\alpha{}_\beta+\lambda^\alpha{}_\kappa\lambda^\kappa{}_\beta.
\ee
On the other hand, the torsion is given by
\be
T^\alpha{}_{\mu\nu}=2\sum_{n=0}^\infty (-1)^n(\lambda^n)^\alpha{}_\beta\partial_{[\mu} \lambda^\beta{}_{\nu]}\simeq 2\Big(\partial_{[\mu} \lambda^\alpha{}_{\nu]}-\lambda^\alpha{}_\rho\partial_{[\mu} \lambda^\rho{}_{\nu]}\Big).
\label{TPalatini}
\ee
We can now also solve for the metric in terms of $\lambda^\alpha{}_\beta$. At zeroth order we obtain that the metric must be constant as it corresponds for the Minkowski background. At first order we have $g_{\mu\nu}\simeq\eta_{\mu\nu}+h_{\mu\nu}$ with $h_{\mu\nu}$ determined by \eqref{eq:Lambdag} to be
\be
h_{\mu\nu}=2\lambda_{(\mu\nu)}+\mathcal{O}(\lambda^2).
\ee
Thus, very much like in the veirbeins approach, the fundamental field $\lambda_{\mu\nu}$ can be decomposed into a symmetric piece that determines the metric perturbation and the complementary antisymmetric part that will encode the remaining components. It is now straightforward to see that the Palatini approach will exactly reproduce all the results obtained above upon the replacement $\lambda_{\mu\nu}\rightarrow A_{\mu\nu}$. We only need to note that the torsion dependence on $\lambda_{\alpha\beta}$ in \eqref{TPalatini} is the same as the one in terms of $A_{\mu\nu}$ above and this will remain at all orders in perturbation theory as can be seen by comparing \eqref{TLAmbda} and \eqref{projT}. Since the Lagrangians are built in terms of the same torsion scalars, both approaches are indeed completely equivalent.

In order to fully show the equivalence of both approaches, we will end this section by obtaining the relation of $\Lambda^\alpha{}_\beta$ and the vierbein. We will use matrix notation to alleviate the notation so that Eq. \eqref{eq:Lambdag} can be expressed as
\be
\hg\,\hL^{-1}\,\partial_\alpha\hL+\partial_\alpha\hL^T\,(\hL^{-1})^T\,\hg=\partial_\alpha\hg.
\label{eq:eqmet2}
\ee
On the other hand, from the relation between the metric and the vierbein $\hg=\he^T\,\hat{\eta}\,\he$ we find
\be
\partial_\alpha\hg=\partial_\alpha\he^T\hat{\eta}\,\he+\he^T\,\hat{\eta}\,\partial_\alpha\he=\partial_\alpha\he^T\big(\he^T\big)^{-1}\,\hg+\hg\,\he^{-1}\,\partial_\alpha\he
\label{eq:eqmet3}
\ee
where we have used that $\hat{\eta}\,\he=\big(\he^T\big)^{-1}\,\hg$. A direct comparison of \eqref{eq:eqmet2} and \eqref{eq:eqmet3} allows to conclude that the matrix $\hL$ plays the role of the vierbein, as one would have expected, and, consequently, the solution for the metric in terms of $\hL$ obtained from \eqref{eq:Lambdag} coincides with the relation between the vierbein and the metric\footnote{Let us emphasise that the choice of $\hat{\eta}$ is conventional with the physical motivation to have Lorentz frames, but any constant internal metric will give equally valid solutions. This originates from the global symmetry $\hat{\Lambda}\rightarrow \hat{U}\hat{\Lambda}$ with $\hat{U}$ a constant element of $GL(4,\mathbb{R})$, that leaves the inertial connection in \eqref{eq:inertialconnection} invariant and is therefore present in Eq. \eqref{eq:eqmet2}.}. Since this is an exact relation, we conclude that the Palatini formulation is completely equivalent to the vierbein approach and we have shown how the equivalence is realised at all orders.

\section{Discussion}
\label{sec:discussion}

In this work we have revisited the number and the nature of the perturbative degrees of freedom of New GR around a flat Minkowski background from different perspectives and extended it to include cubic interactions. The linear theory generally contains a symmetric $h_{\mu\nu}$  and an antisymmetric $b_{\mu\nu}$ fields which mix already in the quadratic Lagrangian. These fields amount to a total of $10+6=16$ dofs, as it corresponds to the number of components of the vierbein $e^a{}_\mu$. Since the theory is Diff invariant, this gauge symmetry removes 8 dofs, so the maximum number of perturbative dofs around Minkowski in New GR is 8. We have explicitly shown how the Diffs act on $h_{\mu\nu}$ and $b_{\mu\nu}$ and obtained the corresponding Bianchi identities as well as their consistency when including couplings to matter. 

After the analysis of the linear theory, we moved to our main goal and explored the cubic Lagrangian. We started by giving the general expression of the cubic interactions. An important point was to obtain how the Diffs invariance is realised at this order. We computed how they are realised directly on the perturbations of the veirbein as well as on the  fields $h_{\mu\nu}$ and $b_{\mu\nu}$. As one would expect, at this order, the realisation of the Diffs on the latter are no longer diagonal and their transformations mix. The corresponding Bianchi identities have also been obtained at this order. In order to study the properties of the cubic Lagrangian and, in particular, whether the gauge symmetry of the 2-form field in the linear theory can be extended to higher orders, we have worked out the decoupling limit of the theory. In that limit, we have explicitly checked the non-trivial relations imposed by the global Lorentz symmetry and we have obtained the main result of this work, namely, that the gauge symmetry of the quadratic Lagrangian cannot be extended to cubic order. An important consequence of this result is that the Minkowski background is strongly coupled due to a discontinuity in the number of dof's. We have concluded the impossibility of deforming the gauge symmetry of the quadratic Lagrangian at cubic order by explicitly showing the non-existence of the would-be Bianchi identities. An alternative manner to obtain our no-go result would be to check if the linear gauge symmetry can be non-linearly deformed into a closed algebra, as it is done for instance in \cite{Wald:1986bj} for the case of spin-1 and spin-2 fields. At the cubic order, this would require to check if there is any choice of $\alpha_i$ in \eqref{deltaomega1} so that the integrability condition imposed by Frobenius theorem $[\delta_{\theta_1},\delta_{\theta_2}]\omega=\delta_\theta\omega$ is satisfied. 

The breaking of the gauge symmetry for the 2-form field sector does not necessarily mean a pathological behaviour as, for instance, in the case of a massive 2-form. The problem for New GR is that the consistency of the linear theory around Minkowski requires the St\"uckelberg fields to enjoy the gauge symmetry because, otherwise, already at that order the theory will contain ghosts. However, as we have shown, this symmetry cannot be maintained once the interactions are included, which means that the gauge symmetry can only arise as an accidental symmetry of the linear theory. Since the interactions will bring in new propagating dof's, these modes will suffer from a severe strong coupling problem around the Minkowski background, thus making such a background ill-defined from a perturbative point of view. This strong coupling problem is a recurrent issue for backgrounds that exhibit some residual gauge symmetry. Within the context of teleparallel theories, the $f(T)$ extensions of the TEGR have been argued to contain one extra-mode with respect to GR \cite{Ferraro:2018tpu}, but the computation of the perturbations around Minkowski and cosmological backgrounds has not revealed any extra dof \cite{Golovnev:2018wbh}, thus suggesting the presence of a strong coupling problem of those backgrounds. This could in turn originate from the existence of some residual local Lorentz symmetries for those backgrounds \cite{Ferraro:2014owa}. For flat geometries with non-metricity but trivial torsion (called symmetric teleparallelisms), it has also been noticed \cite{Jimenez:2019ovq} that the cosmological perturbations of the $f(Q)$ extensions of the corresponding equivalent of GR generically contain two extra scalar modes in the presence of matter fields, but they disappear around maximally symmetric backgrounds. Remarkably, this feature was identified to result from an enhancement of the gauge symmetries of those backgrounds.

In order to illustrate how the problem of having strongly coupled modes comes about, let us imagine a general configuration and a trajectory in phase space approaching the Minkowski solution. As we get closer to the Minkowski fixed point, the kinetic term of some would-be perturbative modes (that could be eventually identified with {\it longitudinal} or {\it gauge} modes of the St\"ueckelberg fields) becomes smaller since we know that they disappear in the exact Minkowski solution where the residual gauge symmetry appears. This in turn means that, if we canonically normalise these modes, their couplings to other modes or to themselves will grow arbitrarily large as we go arbitrarily close to the Minkowski solution and, consequently, they will be strongly coupled. Thus, this solution cannot be perturbatively treated and will be impossible to reach from a general configuration. In other words, this Minkowski solution cannot represent a stable attractor of the system.

This result is not surprising since there was no reason a priori for the theory to exhibit a gauge symmetry at all orders because this would have required the existence of some underlying structure. This underlying structure would not have needed to be apparent and its existence would have required unveiling its origin. Our results were also expected in view of the Hamiltonian analysis performed in the literature \cite{Cheng:1988zg,Blixt:2018znp,Blixt:2019ene}. In particular, the devoted analysis in \cite{Cheng:1988zg} to the family of theories within New GR with $2c_1+c_2+c_3$ revealed that these theories do not contain first class constraints associated to a gauge symmetry. It was found however that they do contain three primary constraints and this was confirmed in \cite{Blixt:2018znp,Blixt:2019ene}, although the constraints algebra was not obtained. The existence of these constraints is not completely apparent in the decoupling limit of the cubic Lagrangian. If we focus on the terms with two derivatives, that contribute to the Hessian, or to the principal part of the field equations or to velocities-dependent terms in the conjugate momenta, we find
\be
\Lag^{(3)}\supset \frac{1}{M}\omega^{0i}\partial_0\omega_{ij}\partial_0 \omega^{0j}
\ee
that give non-vanishing contributions. If we split the 2-form into electric $E_i=\omega_{0i}$ and magnetic $\omega_{ij}=\epsilon_{ijk} B_k$ parts\footnote{These electric and magnetic components of $\omega_{\mu\nu}$ are precisely the S\"uckelberg fields corresponding to boosts and spatial rotations respectively.}, the Hessian can be written as
\bea
\frac{\partial^2\Lag}{\partial \dot{X}_i\partial\dot{X}_j}=
(2a_1-a_2)\left(
\begin{array}{cc}
 0 &\frac{1}{M}\epsilon_{ijk} E_k      \\
 -\frac{1}{M}\epsilon_{ijk} E_k & \delta_{ij}    
\end{array}
\right)
\eea
where $\vec{X}=(\vec{E},\vec{B})$. We see that the cubic interactions introduce non-trivial off-diagonal terms, which are in turn the responsible for the breaking of the first-class constraints of the quadratic Lagrangian. However, if we compute the determinant of the above Hessian, it is straightforward to see that 
\be
\det\left(\frac{\partial^2\Lag}{\partial \dot{X}_i\partial\dot{X}_j}\right)=0.
\ee
The degeneracy of the Hessian indicates the presence of constraints in agreement with the Hamiltonian analyses in \cite{Cheng:1988zg,Okolow:2011np,Blixt:2018znp,Blixt:2019ene}. Let us note that the degenerate Hessian guarantees the existence of at least one constraint, while the Hamiltonian analysis shows that there are three primary constraints. We will not work out the constraints algebra in the decoupling limit here, but we expect to find the same number of constraints. A cautionary comment is in order. The presence of three secondary class constraints is an evidence that the St\"uckelberg fields propagate three additional dof's, as it corresponds to a massive 2-form field so it can be a perfectly healthy and viable theory. It is thus worth emphasising that the main message of our analysis is to explicitly show the existence of strong coupling problems around a physically relevant background such as Minkowski. The root of the problem can then be traced back to the fact that the 2-form becomes massless around Minkowski where it exhibits an accidental gauge symmetry, while only backgrounds where the 2-form remains massive can be free of strong coupling problems. Thus, a natural question is to what extent more general backgrounds can accommodate New GR without strong coupling issues\footnote{One could wonder whether the inclusion of the parity-violating sector, as in \cite{MuellerHoissen:1983vc}, could alleviate the strong coupling problems. Since this sector will also contribute derivative terms in the quadratic action, we do not expect any improvement from it. We thank Tomi S. Kovisito for pointing this out to us.}. We would expect the 2-form to acquire a certain mass at quadratic order determined by the background torsion so that it vanishes as the torsion goes to zero. In this respect, a de Sitter background seems an important background worth studying both from a theoretical and a phenomenological perspective, since it corresponds to a maximally symmetric background and our universe seems to contain a non-vanishing cosmological constant. In a de Sitter background, it would be natural to expect the generation of a mass term for the 2-form parametrically given by the value of the cosmological constant and this could evade the strong coupling of Minkowski. However, the smallness of the cosmological constant will presumably lower the scale at which some interactions will become relevant (that will actually blow up in the strict limit of vanishing cosmological constant, recovering the Minkowski result obtained here) so that a detailed study of such interactions should also be consider. We will leave these important questions for future work.

We will end by noticing that the presence of a pathological 2-form field as an obstruction to have viable modifications of gravity is not particular of New GR, but it is shared by other modified gravity scenarios. For instance, the non-symmetric gravity theories introduced in \cite{Moffat:1978tr} where the metric is allowed to have an antisymmetric part also features a pathological 2-form field, precisely associated to the 2-form that describes the anti-symmetric piece of the metric, as shown in \cite{Damour:1991ru,Damour:1992bt}. There are some notorious similarities with New GR, but also important differences. While obtaining the gauge symmetry at quadratic order in New GR requires a tuning of parameters, the simplest realisation of non-symmetric gravity directly gives a massless Kalb-Ramond field at that order. On the other hand, very much like cubic interactions are irreconcilable with the gauge symmetry in New GR, couplings of the 2-form field to gravity\footnote{By this we mean couplings of the 2-form associated to the antisymmetric part of the metric with the symmetric sector.} in non-symmetric gravity also break the gauge symmetry, resulting in the same discontinuity of dof's that we found for New GR. It was argued that a cosmological constant term providing a mass to the 2-form at quadratic order could resolve the problem, as we also suggested above for New GR. This solution does not seem to completely resolve the issue in non-symmetric gravity however (see e.g. \cite{BeltranJimenez:2019acz}). Another important difference is the origin of the 2-form, being the breaking of local Lorentz symmetry in New GR and the non-symmetric nature of the metric in non-symmetric gravity. Pathological 2-form fields are also found in general higher order curvature theories in the metric-affine formalism \cite{BeltranJimenez:2019acz}. In fact, restricting to theories built in terms of the Ricci tensor alone, the theories can be recast in a form of non-symmetric gravity theories (barring differences in the matter sector couplings), thus sharing the same pathologies, now associated to the non-projective invariant sector of the theories. It is possible to avoid the pathologies by imposing additional symmetries (e.g. a projective symmetry) or additional constraints. These are routes that could also cure some of the New GR Lagrangians and are worth exploring. It is interesting in any case, how pathological 2-forms, which naturally arise from numerous high energy theories like string theories, seem to be at the heart of fundamental obstructions for a variety of consistent modifications of GR.

\acknowledgments
We would like to thank Tomi S. Koivisto and Daniel Blixt for useful discussions. J.B.J. acknowledges support from the  {\it Atracci\'on del Talento Cient\'ifico en Salamanca} programme and the MINECO's projects FIS2014-52837-P and FIS2016-78859-P (AEI/FEDER). This article is based upon work from CANTATA COST (European Cooperation in Science and Technology) action CA15117, EU Framework Programme Horizon 2020.

\end{document}